\documentclass[11pt, a4paper, logo, copyright]{googlecloud}

\pdftrailerid{redacted}

\makeatletter
\renewcommand\bibentry[1]{\nocitep{#1}{\frenchspacing\@nameuse{BR@r@#1\@extra@b@citeb}}}
\makeatother

\usepackage[authoryear, sort&compress, round]{natbib}
\usepackage{latexsym}
\usepackage{nicefrac}
\usepackage{multirow}
\usepackage{float}
\usepackage{subcaption}   % replaces the old `subfigure` package (incompatible with caption.sty)

\title{ERA: Embed, Route, Adapt for Cold-Start Cross-Model Latent Retrieval}

\renewcommand{\today}{}

\author[1]{\fontsize{10.0pt}{10.0pt}\selectfont Venkata Chivukula}
\author[1]{\fontsize{10.0pt}{10.0pt}\selectfont Zifeng Wang}
\author[1]{\fontsize{10.0pt}{10.0pt}\selectfont Jinsung Yoon}
\affil[1]{\fontsize{9.0pt}{9.0pt}\selectfont Google}

\begin{document}

\begin{abstract}
Production retrieval systems are tied to the embedding model that indexed
their corpus. Moving queries to a different encoder normally requires
re-embedding the whole corpus, and at cold start there are no query logs while
every document embedded by the new encoder incurs an API or compute cost.
We present ERA (Embed, Route, Adapt), a system for cold-start cross-model
latent retrieval.
ERA spends a small anchor budget on maximally informative documents, selected
from the existing target index via Cosine Farthest Point Sampling, trains a lightweight adapter with a
Manifold Alignment Loss that couples directional alignment with orthogonal
regularization to prevent representation collapse, and pairs the adapter with a
deterministic router that escalates the rare, hard queries that define
perceived retrieval quality.
Across five BEIR datasets, ERA matches or outperforms point-wise $\ell_2$ and
embedding-conversion baselines, with relative gains of up to
35\% nDCG@10 on Arguana, and on clustered corpora its anchor selection also improves prior methods
under a fixed anchor budget.
With 1.2\,ms of adapter and router overhead on top of a $\sim$20\,ms local
encoder, ERA at a 20\% nominal escalation budget reduces
end-to-end query latency from roughly 160\,ms to 38--86\,ms while concentrating
escalations on the queries where the projection fails most.
\end{abstract}

\maketitle

\section{Introduction}
\label{sec:intro}

Retrieval-augmented generation (RAG) grounds language models in
organization-specific knowledge by retrieving from a dense vector index
\citep{lewis2020retrieval, karpukhin2020dense}. Such an index lives in the
latent space of the embedding model that built it: queries must be embedded
by that same model, or inner-product search no longer returns meaningful
neighbours. In practice, every query must therefore go through the indexing
model, whether that is an API with per-call cost and network latency or a
large self-hosted model with its own serving footprint. Using a different
encoder for queries normally requires re-embedding the entire corpus. That
can be expensive, and in many deployments it is not possible at all: the
index may be managed by a third party, the raw documents may no longer be
accessible, or the corpus may be ingested continuously.

We study \emph{cross-model latent retrieval}: a query is encoded by one
model, projected into the latent space of a different model, and retrieved
directly against that model's existing index. Either model may be
open-weight or served through an API. Prior work has shown that mappings
between embedding spaces are feasible, from unsupervised alignment
\citep{jha2024universal, dar2025minivec2vec} to embedding conversion that
serves a smaller model online while retaining a larger model offline
\citep{yoon2025embedding} and effective point-wise distillation
\citep{vujanic2025leaf}. We build on these insights and focus on the setting
that arises at the very start of every deployment: the \emph{cold start}.

At cold start, three constraints hold at once. First, there are no query
logs and no relevance labels. Second, the target index already exists, but every
anchor document must also be embedded by the source model, in API calls or in
compute, so the training budget is measured in anchors. Third, a projection will not be perfect, so some
queries must still be answered by the target model, and the queries that
matter most are rarely the common ones. Real traffic is dominated by easy
queries, while the comparatively rare complex queries are where retrieval
quality is most visible to users and where a projection is most likely to
fail. A cold-start system must therefore decide \emph{which documents} to
spend its anchor budget on, \emph{how} to learn a faithful projection from
very few anchors, and \emph{which queries} still deserve the target model.

\begin{figure*}[t]
\centering
\includegraphics[width=0.90\textwidth]{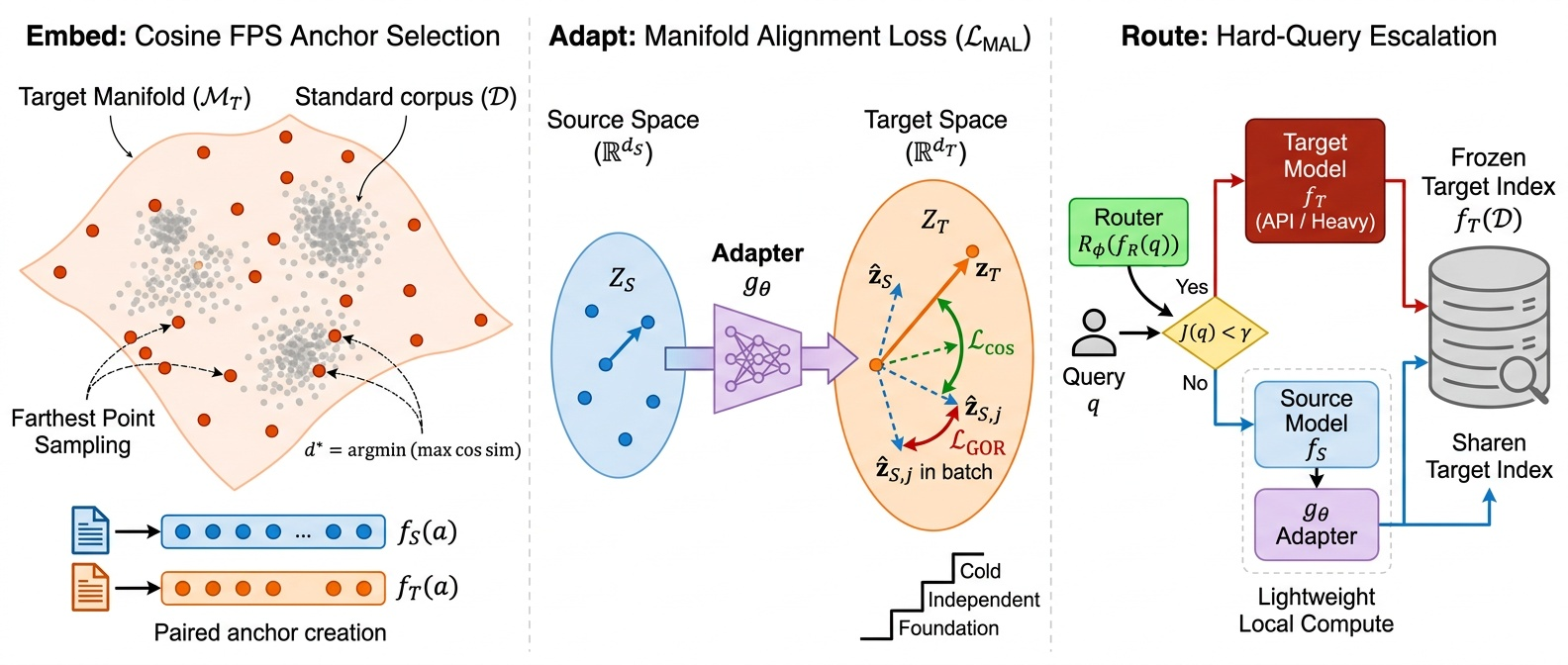}
\caption{Overview of the ERA (Embed, Route, Adapt) architecture for cold-start
cross-model latent retrieval.}
\label{fig:overview}
\end{figure*}

We present \textbf{ERA (Embed, Route, Adapt)}\footnote{Not to be confused with the
Efficient Retrieval Adapter of \citet{maekawa2026align}, which requires labeled
query--document pairs.}, an end-to-end system for
cold-start cross-model latent retrieval (Figure~\ref{fig:overview}).
\textbf{Embed:} ERA selects anchor documents with Cosine Farthest Point
Sampling (FPS) in the target space \citep{qi2017pointnet}. Real corpora are
highly redundant, and uniform sampling spends most of a small budget on
near-duplicate regions; FPS covers the distinct regions of the target
manifold before revisiting any of them.
\textbf{Adapt:} ERA trains a lightweight adapter with a Manifold Alignment
Loss that couples directional alignment with Global Orthogonal
Regularization. The orthogonality term spreads projections across the
target hypersphere and prevents the representation collapse
\citep{gao2019representation} that point-wise objectives are prone to when
anchors are scarce.
\textbf{Route:} ERA pairs the adapter with an approximate router that
escalates the queries most likely to be mishandled by the projection to the
target model. The router runs on any lightweight encoder (open-weight in our
implementation, for latency) that is separate from the escalation target,
so escalated queries never pay for two expensive encoders. As unlabeled
query logs become available, the same pipeline continues to improve through
in-domain and cross-domain query adaptation.

We evaluate ERA on five BEIR datasets \citep{thakur2021beir}, projecting
both an open-weight (\texttt{Qwen-Embedding-3-0.6B}) and an API-served
(\texttt{text-embedding-004}) query encoder into the
\texttt{gemini-embedding-001} index, and a reverse-direction control
(Gemini $\rightarrow$ Qwen) confirms that ERA faithfully reproduces whichever
target space it is aligned to.
ERA matches or outperforms MSE-based and
embedding-conversion baselines, improving nDCG@10 by up to 35\% (relative) on
Arguana and 12\% on FiQA, and on clustered corpora FPS improves even prior methods at a fixed anchor
budget. At a 20\% escalation budget, the deterministic router escalates
37.5\% of the huge-gap queries on SciFact and 32.0\% on SciDocs, whereas a
random policy at the same budget misses every huge-gap SciFact query in
roughly one deployment out of six. With 1.2\,ms of adapter and router overhead on top of a $\sim$20\,ms local
encoder, end-to-end latency
drops from roughly 160\,ms to 38--86\,ms.

In summary, we make the following contributions:
\begin{itemize}
\item We formalize \emph{cold-start cross-model latent retrieval}: label-free
      projection of queries from one model into the existing index of
      another under an anchor budget, without re-indexing.
\item We show that budget-aware anchor selection via Cosine FPS is an
      important and loss-independent design choice, whose benefit grows with
      corpus redundancy.
\item We introduce a Manifold Alignment Loss that couples directional
      alignment with orthogonal regularization, preventing representation
      collapse in the data-scarce regime.
\item We design an approximate escalation router that targets the rare,
      hard queries that define perceived quality, at negligible local cost.
\item To our knowledge, ERA is the first end-to-end system that combines
      label-free cold-start alignment, budget-aware anchor selection, and
      hard-query escalation for zero-reindex cross-model retrieval.
\end{itemize}

\section{Related Work}
\label{sec:related}

\paragraph{Retrieval-augmented generation and index coupling.}
Retrieval-augmented generation grounds language models in external knowledge
by retrieving passages from a dense vector index
\citep{lewis2020retrieval, karpukhin2020dense, gao2023retrieval}. Because
nearest-neighbour search in modern vector databases relies on inner products
within a fixed latent space \citep{malkov2018efficient, johnson2019billion},
the query encoder and the corpus index must live in the same space. Changing
the query encoder normally requires re-embedding the corpus, which motivates
methods that project embeddings from one model into the space of another.

\paragraph{Cross-model embedding alignment and conversion.}
Classical knowledge distillation aligns a student to a teacher through internal
representations or logits \citep{hinton2015distilling, wang2020minilm}. When
models are accessed only through their output embeddings, alignment must
operate directly on the two latent spaces. While \textit{vec2vec}
\citep{jha2024universal} and \textit{mini-vec2vec} \citep{dar2025minivec2vec}
showed that unpaired embedding clouds can be aligned via adversarial training
or Procrustes analysis \citep{schonemann1966generalized}, practical retrieval
migration embeds a shared set of reference texts with both models. In this
paired setting, \textit{Embedding-\allowbreak Converter}
\citep{yoon2025embedding} learns a post-hoc mapper combining point-wise
regression with global and local similarity preservation, establishing the
paradigm our work builds on most directly: encoding queries with a lightweight
model online while searching an index built by a stronger model offline. In a
related direction, \textit{LEAF} \citep{vujanic2025leaf} shows that
point-wise $\ell_2$ alignment is a strong baseline for teacher-aligned
embedding distillation, and \textit{QuARI} \citep{xing2025quari} adapts query
representations to improve retrieval. Closest to our setting,
\textit{Drift-Adapter} \citep{vejendla2025drift} maps queries from an upgraded
model into a legacy index with Procrustes, low-rank affine, or residual MLP
adapters, and \citet{maekawa2026align} learn a query-side adapter (also named
ERA) via unsupervised alignment followed by few-shot supervised adaptation; our
ERA is fully label-free and adds budget-aware anchor selection and escalation
routing.

ERA builds directly on Embedding-Converter and LEAF and focuses on the
\emph{cold-start} regime. Both prior methods sample reference texts uniformly
at random, which is natural when paired embeddings are plentiful. At cold
start, however, every anchor must be embedded by the source model, and real corpora are
highly redundant: in a corpus of 1{,}000 documents where 900 are
near-duplicates of one topic and 100 span distinct topics, uniform sampling of
100 anchors spends roughly 90 of them on the redundant cluster and covers only
about 10 distinct regions of the target space. ERA instead selects anchors with
Cosine Farthest Point Sampling \citep{qi2017pointnet} so that the budget covers
distinct regions first---a design choice that, on clustered corpora, improves both Embedding-Converter
and LEAF as well as ERA under the same anchor budget.

\paragraph{Representation geometry and adaptive routing.}
Learned embedding spaces are known to be anisotropic, with vectors concentrating
in a narrow cone on the hypersphere
\citep{gao2019representation, ethayarajh2019how}. Contrastive and metric
learning address this by encouraging uniformity across the sphere, including
Global Orthogonal Regularization \citep{zhang2017learning} and related objectives
\citep{wang2020understanding, chen2020simple, vera2025embeddinggemma}. When a
projection is learned from only a small anchor set, point-wise objectives can
collapse toward a low-variance subspace; ERA pairs directional alignment with
orthogonal regularization to preserve angular spread in this regime. Finally,
confidence-based routing between a fast model and a stronger fallback has been
studied for language model generation \citep{ong2025routellm}. ERA brings
budget-controlled escalation to cross-model retrieval, targeting the rare,
complex queries where a low-dimensional projection falls short of the target
space.

\section{Methodology}
\label{sec:methodology}

\subsection{Problem Formulation: Cross-Model Latent Retrieval}
\label{subsec:formulation}

Let $\mathcal{D}$ be a document corpus indexed by a target embedding model
$f_T : \mathcal{D} \to \mathbb{R}^{d_T}$, producing a frozen vector index
$f_T(\mathcal{D}) \subset \mathcal{M}_T$. At query time we wish to encode incoming
queries with a distinct source model $f_S : \mathcal{Q} \to \mathbb{R}^{d_S}$
(typically with $d_S \neq d_T$), where either $f_S$ or $f_T$ may be open-weight
or served behind an API. Because nearest-neighbour retrieval relies on inner
products within $\mathcal{M}_T$, the two spaces cannot be queried across directly.
Our goal is to learn a lightweight, non-linear projection
$g_\theta : \mathbb{R}^{d_S} \to \mathbb{R}^{d_T}$ such that inner-product
rankings in the target index are preserved for any query $q$ and document $d$:
\begin{equation}
    \langle g_\theta(f_S(q)), f_T(d) \rangle \approx \langle f_T(q), f_T(d) \rangle.
\end{equation}
This allows queries encoded by $f_S$ to search the existing index $f_T(\mathcal{D})$
directly without re-embedding the corpus.

\subsection{Embed: Cold-Start Anchor Selection via Cosine FPS}
\label{subsec:fps}

At cold start, no query logs or relevance labels exist. The target embeddings
$f_T(\mathcal{D})$ are already stored in the index, so anchor selection needs
only vector operations over them; the cost lies in embedding each selected
anchor with $f_S$, which incurs an API or compute cost. Prior
conversion methods sample reference texts uniformly at random from $\mathcal{D}$
\citep{yoon2025embedding, vujanic2025leaf}. However, real corpora exhibit heavy
topical clustering: if 90\% of a corpus belongs to a single dense topic and
10\% spans distinct tail topics, a uniform sample of $N$ anchors spends roughly
$0.9N$ of its budget on near-duplicate vectors while leaving the rest of the
target manifold unmapped.

To maximize geometric coverage per anchor, we select the anchor set
$\mathcal{A} \subset \mathcal{D}$ ($|\mathcal{A}| = N$) using Cosine Farthest
Point Sampling (FPS) \citep{qi2017pointnet}. Starting from an initial document
$d_0$, we iteratively add the document whose embedding has the smallest maximum
cosine similarity to the anchors already selected:
\begin{equation}
    d^* = \arg\min_{d \in \mathcal{D} \setminus \mathcal{A}}
    \left( \max_{a \in \mathcal{A}}
    \frac{f_T(d)^\top f_T(a)}{\|f_T(d)\| \, \|f_T(a)\|} \right).
\end{equation}
Embedding $\mathcal{A}$ with $f_S$ and pairing each anchor with its stored index
vector $f_T(a)$ yields a diverse set of paired
anchors $\{(f_S(a), f_T(a))\}_{a \in \mathcal{A}}$ that spans the distinct
regions of $\mathcal{M}_T$ before revisiting dense clusters.

\subsection{Adapt: Manifold Alignment Loss and Curriculum}
\label{subsec:adapt}

\paragraph{Manifold Alignment Loss ($\mathcal{L}_{\mathrm{MAL}}$).}
Learning $g_\theta : \mathbb{R}^{d_S} \to \mathbb{R}^{d_T}$ from a small anchor
set is under-constrained under point-wise $\ell_2$ regression: to minimize
average error, the adapter can map vectors into a narrow, low-variance cone on
the target hypersphere \citep{gao2019representation}, which degrades
nearest-neighbour discrimination \citep{johnson2019billion, malkov2018efficient}.
Let $Z_S \in \mathbb{R}^{B \times d_S}$ and $Z_T \in \mathbb{R}^{B \times d_T}$
denote a mini-batch of source and target embeddings, and let
$\hat{Z}_S = g_\theta(Z_S)$ be the projected batch. We align individual
directions with a cosine loss,
\begin{equation}
    \mathcal{L}_{\cos} = \frac{1}{B} \sum_{i=1}^{B}
    \left( 1 - \frac{\hat{z}_{S,i}^\top z_{T,i}}
    {\|\hat{z}_{S,i}\| \, \|z_{T,i}\|} \right),
\end{equation}
and prevent angular collapse across the batch with Global Orthogonal
Regularization (GOR) \citep{zhang2017learning, wang2020understanding, chen2020simple}, which
penalizes off-diagonal cosine correlations among projected vectors:
\begin{equation}
    \mathcal{L}_{\mathrm{GOR}} = \frac{1}{B(B - 1)} \sum_{i \neq j}
    \left( \frac{\hat{z}_{S,i}^\top \hat{z}_{S,j}}
    {\|\hat{z}_{S,i}\| \, \|\hat{z}_{S,j}\|} \right)^2.
\end{equation}
The alignment objective combines the two terms,
\begin{equation}
    \mathcal{L}_{\mathrm{MAL}} = \mathcal{L}_{\cos} + \mathcal{L}_{\mathrm{GOR}},
\end{equation}
which couples point-wise directional accuracy with hyperspherical uniformity.
In practice, each term is scaled by a learned homoscedastic uncertainty weight
\citep{kendall2018multi}, which performed on par with fixed equal weights in
our ablations, so no per-dataset weight tuning is required.

\paragraph{Three-stage adaptation curriculum.}
ERA supports three progressive, label-free deployment stages as unlabeled text
becomes available:
(1)~\textbf{Cold} trains $g_\theta$ solely on the $N$ document anchors selected
by Cosine FPS, requiring zero queries;
(2)~\textbf{Independent} warm-starts from the Cold adapter and continues
training on unlabeled in-domain queries embedded by $f_S$ and $f_T$; and
(3)~\textbf{Foundation} pools unlabeled queries from heterogeneous source
domains to learn a general query-to-document projection geometry that transfers
zero-shot to unseen English retrieval domains.

\subsection{Route: Hard-Query Escalation}
\label{subsec:routing}

When $d_S < d_T$, an asymmetric projection has finite capacity, and complex
queries can suffer large retrieval gaps relative to native target encoding.
In production traffic, the vast majority of queries are simple and well-handled
by $g_\theta(f_S(q))$, whereas a small tail of complex queries determines
perceived system quality. However, if $f_S$ is itself a commercial API, running
$f_S(q)$ first and then escalating to $f_T(q)$ would charge two API calls on the
same query.

To avoid paying for two encoders on escalated queries, ERA decouples routing
from $f_S$ by using a fast, lightweight feature extractor $f_R$ (an open-weight
encoder in our implementation for minimal latency, though any inexpensive
encoder suffices). A lightweight router MLP $R_\phi$ takes $f_R(q)$ and predicts
the top-$K$ retrieval agreement between the source and target spaces:
\begin{equation}
    J(q) = \frac{|\mathcal{R}_K(q, f_T) \cap \mathcal{R}_K(q, f_S)|}{K},
\end{equation}
where $\mathcal{R}_K(q, f)$ is the set of top-$K$ documents retrieved for $q$
under encoder $f$. Low agreement marks regions that the source encoder
represents poorly and that the projection is therefore likely to mishandle
(in the run behind Table~\ref{tab:router_gaps}, the first of two training
stages instead uses the frozen projection $g_\theta \circ f_S$ in place of
$f_S$). Because no queries exist at cold start, we train $R_\phi$ on
anchor documents $\mathcal{A}_{\mathrm{r}}$ used as pseudo-queries, retrieving
within the anchor set, via mean squared error:
\begin{equation}
    \mathcal{L}_{\mathrm{router}} = \frac{1}{|\mathcal{A}_{\mathrm{r}}|}
    \sum_{a \in \mathcal{A}_{\mathrm{r}}}
    \left( R_\phi(f_R(a)) - J(a) \right)^2.
\end{equation}
Given an escalation budget (e.g., 20\% of traffic), we set the decision
threshold $\gamma$ to the corresponding quantile of predicted scores
$\hat{J}(q) = R_\phi(f_R(q))$ on held-out calibration queries (or held-out
anchors for datasets without training queries). At inference time,
queries with $\hat{J}(q) < \gamma$ are routed directly to $f_T(q)$; all other
queries are served via $g_\theta(f_S(q))$, yielding a deterministic escalation
policy with negligible local overhead.

\section{Experimental Evaluation}
\label{sec:evaluation}

\subsection{Experimental Setup}
\label{subsec:setup}

\paragraph{Datasets and metrics.}
We evaluate on five BEIR retrieval benchmarks \citep{thakur2021beir} spanning
scientific, financial, biomedical, and argumentative domains: SciFact, FiQA,
NFCorpus, Arguana, and SciDocs. We report ground-truth \textrm{nDCG@10} against
human relevance judgments and \textrm{T-nDCG@10} (topological fidelity, treating
the target encoder's top-10 retrieved list as ground truth). We evaluate on the
test queries with at least one relevant passage in the TFDS BEIR release
(SciFact 247, NFCorpus 185, ArguAna 1{,}401, SciDocs 994, FiQA 648). All reported
uncertainties ($\pm$) denote one standard deviation across three random seed
initializations.

\paragraph{Model configurations.}
We use \texttt{gemini-embedding-001} ($d_T = 3072$) as the target index encoder
$f_T$. We evaluate two source query encoders $f_S$: an open-weight model
(\texttt{Qwen-Embedding-3-0.6B}, $d_S = 1024$) and a commercial API model
(\texttt{text-embedding-004}, $d_S = 768$). We also run a reverse-direction
negative control projecting \texttt{gemini-embedding-001} ($3072$-D) into a
\texttt{Qwen-Embedding-3-0.6B} ($1024$-D) index. Both the adapter $g_\theta$
and the router $R_\phi$ are 2-layer MLPs with SwiGLU activations
\citep{shazeer2020glu} and hidden dimension $4096$.

\paragraph{Training details and baselines.}
Unless varied in scaling ablations, the Cold phase uses $|\mathcal{A}| = 4000$
document anchors drawn uniformly at random (Cosine FPS is evaluated separately
in Section~\ref{subsec:anchor_efficiency}) and trains for under 10 minutes on a single
TPUv3 node using Adam \citep{kingma2015adam} with learning rate $10^{-4}$. The
Independent phase uses only the official training query split of each dataset
(the validation split where one exists, with 10\% held out for router calibration)
(zero test-query leakage), and the Foundation phase pools the training queries
of SciFact, NFCorpus, and FiQA; Arguana and SciDocs, which provide no training
queries, are therefore evaluated zero-shot.
The Cold adapter and router use only anchor documents. To set the escalation
threshold, we use the router-encoder embeddings of 10\% held-out unlabeled
queries when a dataset provides them (no labels or target-model calls);
ArguAna and SciDocs, which have none, calibrate on held-out anchors.
We compare ERA against two paired post-hoc conversion baselines trained under
the same data (LEAF uses the ERA adapter architecture and
Embedding-Converter a ReLU MLP of the same width): \textbf{Embedding-Converter}
\citep{yoon2025embedding} and \textbf{LEAF} (point-wise $\ell_2$ regression)
\citep{vujanic2025leaf}.
Our objective evolved during development: system-level results
(Tables~\ref{tab:phase_summary}, \ref{tab:router_gaps}, and \ref{tab:latency},
and the appendix) use the listwise-KL-augmented variant
$\mathcal{L}_{\cos}+\mathcal{L}_{\mathrm{KL}}+\mathcal{L}_{\mathrm{GOR}}$ (the
\texttt{Qwen-Embedding-3-0.6B} Cold and Independent columns additionally include
the bilinear term and an anchor-only refinement stage), which
Table~\ref{tab:loss_ablation} shows to be within $0.02$ nDCG@10 of
$\mathcal{L}_{\mathrm{MAL}}$.

\begin{table*}[t]
\caption{Baseline and loss ablation across all five BEIR benchmarks
(\texttt{text-embedding-004} $\to$ \texttt{gemini-embedding-001}, 0\% routing,
after the Cold and Foundation phases),
reporting ground-truth retrieval accuracy (nDCG@10) and topological fidelity to
the target index (T-nDCG@10) with standard deviations over three seeds.
LEAF minimizes the $\ell_2$ error between unit-normalized embeddings, which
coincides with cosine regression up to scale, and thus closely corresponds to
ERA without $\mathcal{L}_{\mathrm{GOR}}$. ERA$^{+}$ denotes a
four-term variant that additionally includes a listwise KL distillation term
over the target's top-100 documents and a signed bilinear alignment term; its
leave-one-out rows isolate each ERA term within the richer objective.}
\label{tab:loss_ablation}
\centering
\small
\setlength{\tabcolsep}{4pt}
\resizebox{\textwidth}{!}{%
\begin{tabular}{@{}llccccc@{}}
\toprule
\textbf{Metric} & \textbf{Method} & \textbf{Arguana} & \textbf{FiQA} & \textbf{NFCorpus} & \textbf{SciDocs} & \textbf{SciFact} \\
\midrule
nDCG@10
& LEAF (Pure $L_2$)                    & 0.340$\pm$0.007 & 0.424$\pm$0.002 & 0.230$\pm$0.008 & 0.159$\pm$0.007 & 0.864$\pm$0.005 \\
& ERA$^{+}$ w/o $\mathcal{L}_{\cos}$    & 0.429$\pm$0.004 & 0.397$\pm$0.011 & 0.226$\pm$0.005 & 0.139$\pm$0.004 & 0.845$\pm$0.004 \\
& ERA$^{+}$ w/o $\mathcal{L}_{\mathrm{GOR}}$ & 0.377$\pm$0.032 & 0.422$\pm$0.004 & 0.233$\pm$0.009 & 0.160$\pm$0.005 & \textbf{0.866$\pm$0.003} \\
& \textbf{ERA ($\mathcal{L}_{\cos}+\mathcal{L}_{\mathrm{GOR}}$)} & \textbf{0.459$\pm$0.017} & \textbf{0.475$\pm$0.006} & \textbf{0.243$\pm$0.002} & \textbf{0.164$\pm$0.004} & 0.865$\pm$0.001 \\
\midrule
T-nDCG@10
& LEAF (Pure $L_2$)                    & 0.664$\pm$0.011 & 0.583$\pm$0.003 & 0.651$\pm$0.007 & 0.504$\pm$0.004 & 0.764$\pm$0.005 \\
& ERA$^{+}$ w/o $\mathcal{L}_{\cos}$    & 0.666$\pm$0.007 & 0.494$\pm$0.006 & 0.589$\pm$0.006 & 0.410$\pm$0.004 & 0.742$\pm$0.005 \\
& ERA$^{+}$ w/o $\mathcal{L}_{\mathrm{GOR}}$ & 0.678$\pm$0.012 & 0.582$\pm$0.007 & 0.658$\pm$0.008 & 0.503$\pm$0.004 & 0.765$\pm$0.004 \\
& \textbf{ERA ($\mathcal{L}_{\cos}+\mathcal{L}_{\mathrm{GOR}}$)} & \textbf{0.742$\pm$0.009} & \textbf{0.604$\pm$0.004} & \textbf{0.662$\pm$0.005} & \textbf{0.534$\pm$0.004} & \textbf{0.771$\pm$0.002} \\
\bottomrule
\end{tabular}}
\end{table*}
\begin{table}[tb]
\caption{Adaptation progression across Cold, Independent, and Foundation phases
(nDCG@10, \texttt{Qwen-0.6B} $\to$ \texttt{gemini-embedding-001}) at 0\% and a
nominal 20\% escalation budget. Parentheses give the observed fraction of test
queries escalated (Foundation phase; it varies by less than one point across
phases, see Table~\ref{tab:unified_absolute_performance}).}
\label{tab:phase_summary}
\centering
\small
\setlength{\tabcolsep}{2pt}
\begin{tabular*}{0.75\textwidth}{@{\extracolsep{\fill}}llccc@{}}
\toprule
\textbf{Dataset} & \textbf{Budget (routed)} & \textbf{Cold} & \textbf{Indep.} & \textbf{Found.} \\
\midrule
SciFact  & 0\%  & 0.676$\pm$.006 & 0.837$\pm$.003 & \textbf{0.840$\pm$.006} \\
         & 20\% (21\%) & 0.733$\pm$.004 & 0.857$\pm$.004 & \textbf{0.860$\pm$.001} \\
\midrule
FiQA     & 0\%  & 0.289$\pm$.000 & 0.354$\pm$.003 & \textbf{0.458$\pm$.004} \\
         & 20\% (10\%) & 0.325$\pm$.004 & 0.376$\pm$.003 & \textbf{0.473$\pm$.004} \\
\midrule
NFCorpus & 0\%  & 0.211$\pm$.004 & 0.224$\pm$.006 & \textbf{0.228$\pm$.008} \\
         & 20\% (25\%) & 0.231$\pm$.003 & 0.241$\pm$.006 & \textbf{0.247$\pm$.007} \\
\midrule
Arguana  & 0\%  & 0.312$\pm$.003 & 0.306$\pm$.006 & \textbf{0.559$\pm$.007} \\
         & 20\% (40\%) & 0.449$\pm$.001 & 0.443$\pm$.002 & \textbf{0.595$\pm$.005} \\
\midrule
SciDocs  & 0\%  & 0.173$\pm$.001 & 0.175$\pm$.001 & \textbf{0.180$\pm$.002} \\
         & 20\% (38\%) & 0.203$\pm$.001 & 0.205$\pm$.001 & \textbf{0.207$\pm$.002} \\
\bottomrule
\end{tabular*}
\end{table}

\subsection{Alignment Accuracy and Loss Ablations}
\label{subsec:cold_start_results}

Table~\ref{tab:loss_ablation} evaluates ERA ($\mathcal{L}_{\cos} +
\mathcal{L}_{\mathrm{GOR}}$) against point-wise $\ell_2$ regression (LEAF) and
single-term ablations at $0\%$ escalation (pure adapter) across all five BEIR
datasets. Two findings stand out. First, ERA outperforms point-wise $\ell_2$
(LEAF) on every dataset in topological fidelity (T-nDCG@10)---with gains of
$+11.7\%$ on Arguana ($0.742$ vs.\ $0.664$), $+5.9\%$ on SciDocs ($0.534$ vs.\
$0.504$), and $+3.6\%$ on FiQA ($0.604$ vs.\ $0.583$)---and beats or matches
LEAF on all five datasets in ground-truth nDCG@10, improving Arguana by
$+35.0\%$ ($0.459$ vs.\ $0.340$), FiQA by $+11.8\%$ ($0.475$ vs.\ $0.424$), and
NFCorpus by $+5.4\%$ ($0.243$ vs.\ $0.230$). Second, the leave-one-out rows of
the four-term variant ERA$^{+}$ confirm that both terms in
$\mathcal{L}_{\mathrm{MAL}}$ are required: removing
directional alignment ($\text{w/o }\mathcal{L}_{\cos}$) degrades T-nDCG@10
sharply across FiQA ($0.604 \to 0.494$), NFCorpus ($0.662 \to 0.589$), and
SciDocs ($0.534 \to 0.410$), while removing orthogonal regularization
($\text{w/o }\mathcal{L}_{\mathrm{GOR}}$) allows projected vectors to collapse
into a narrow cone, dropping Arguana nDCG@10 from $0.459$ to $0.377$ and FiQA
from $0.475$ to $0.422$.

\subsection{Anchor Efficiency: Cosine FPS vs.\ Uniform Sampling}
\label{subsec:anchor_efficiency}

\begin{table}[tb]
\caption{Cold-start anchor efficiency (T-nDCG@10) under Cosine FPS vs.\ uniform
random sampling at anchor budgets $N \in \{100, 400, 1000\}$
(\texttt{Qwen-Embedding-3-0.6B} $\to$ \texttt{gemini-embedding-001}, Cold phase,
single seed; ERA uses the four-term ERA$^{+}$ objective).}
\label{tab:data_efficiency}
\centering
\small
\setlength{\tabcolsep}{2.5pt}
\begin{tabular*}{0.75\textwidth}{@{\extracolsep{\fill}}lllccc@{}}
\toprule
\textbf{Dataset} & \textbf{Model} & \textbf{Samp.} & $\mathbf{N{=}100}$ & $\mathbf{N{=}400}$ & $\mathbf{N{=}1000}$ \\
\midrule
SciFact  & ERA         & FPS  & \textbf{0.190} & \textbf{0.434} & \textbf{0.544} \\
         &             & Rand & 0.175 & 0.401 & 0.509 \\
         & Embed-Conv  & FPS  & 0.051 & 0.300 & 0.457 \\
         &             & Rand & 0.060 & 0.261 & 0.419 \\
         & LEAF        & FPS  & 0.139 & 0.412 & 0.538 \\
         &             & Rand & 0.137 & 0.383 & 0.502 \\
\midrule
NFCorpus & ERA         & FPS  & 0.152 & \textbf{0.401} & \textbf{0.496} \\
         &             & Rand & \textbf{0.188} & 0.379 & 0.478 \\
         & Embed-Conv  & FPS  & 0.046 & 0.296 & 0.412 \\
         &             & Rand & 0.065 & 0.247 & 0.393 \\
         & LEAF        & FPS  & 0.123 & 0.392 & 0.493 \\
         &             & Rand & 0.151 & 0.386 & 0.482 \\
\midrule
Arguana  & ERA         & FPS  & \textbf{0.241} & 0.568 & 0.670 \\
         &             & Rand & 0.142 & 0.361 & 0.563 \\
         & Embed-Conv  & FPS  & 0.091 & 0.413 & 0.550 \\
         &             & Rand & 0.033 & 0.173 & 0.375 \\
         & LEAF        & FPS  & 0.227 & \textbf{0.580} & \textbf{0.672} \\
         &             & Rand & 0.117 & 0.331 & 0.521 \\
\midrule
SciDocs  & ERA         & FPS  & 0.063 & 0.250 & 0.391 \\
         &             & Rand & \textbf{0.090} & \textbf{0.298} & \textbf{0.424} \\
         & Embed-Conv  & FPS  & 0.010 & 0.119 & 0.265 \\
         &             & Rand & 0.022 & 0.149 & 0.302 \\
         & LEAF        & FPS  & 0.031 & 0.200 & 0.373 \\
         &             & Rand & 0.061 & 0.265 & 0.408 \\
\midrule
FiQA     & ERA         & FPS  & 0.036 & 0.173 & \textbf{0.317} \\
         &             & Rand & \textbf{0.041} & \textbf{0.183} & 0.302 \\
         & Embed-Conv  & FPS  & 0.004 & 0.068 & 0.212 \\
         &             & Rand & 0.007 & 0.086 & 0.212 \\
         & LEAF        & FPS  & 0.017 & 0.124 & 0.270 \\
         &             & Rand & 0.021 & 0.146 & 0.270 \\
\bottomrule
\end{tabular*}
\end{table}

\begin{figure*}[t]
\centering
\begin{subfigure}[t]{0.48\textwidth}
  \centering
  \includegraphics[width=\linewidth]{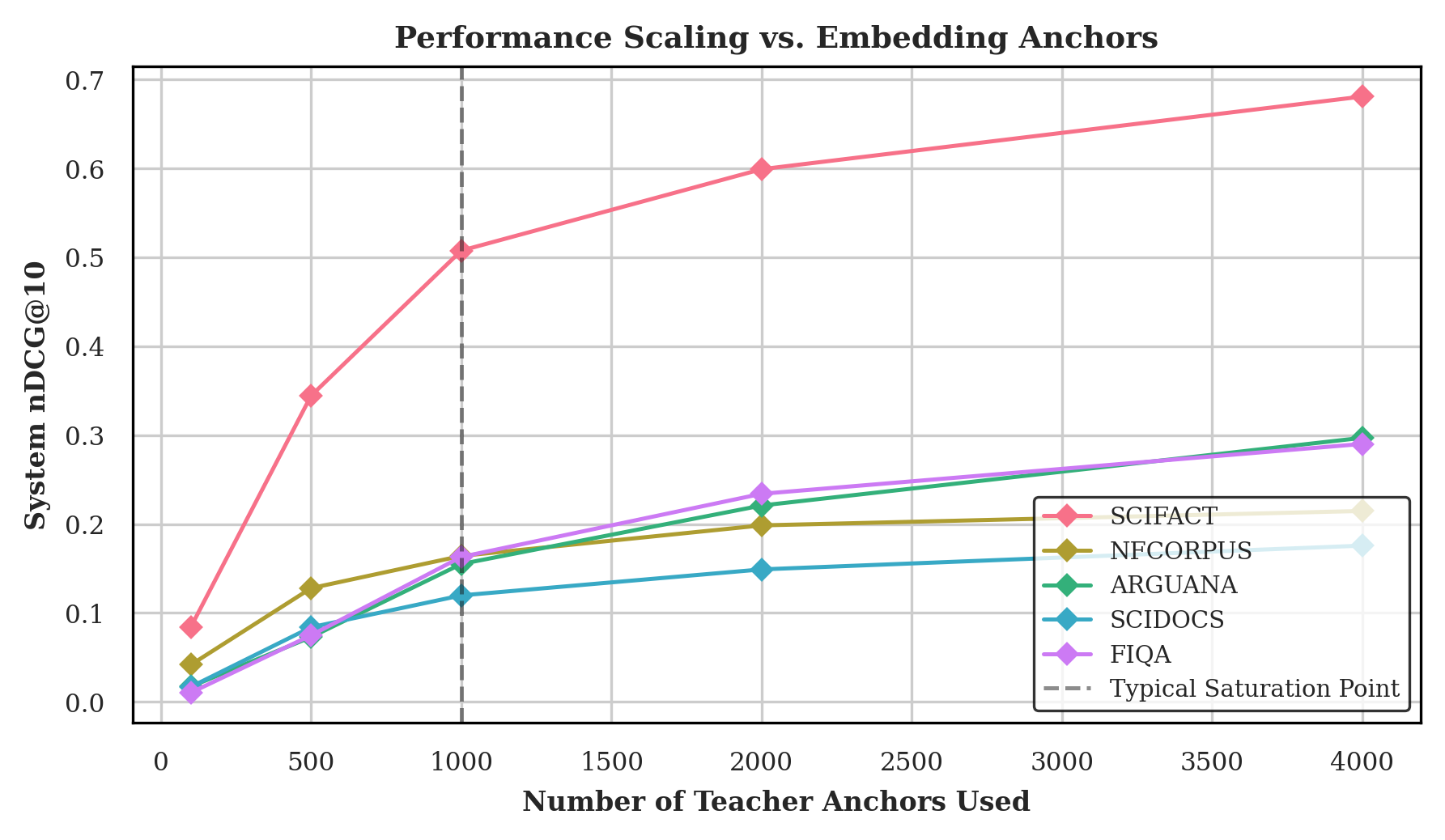}
  \caption{\texttt{Qwen-Embedding-3-0.6B} $\to$ \texttt{gemini-embedding-001}}
\end{subfigure}\hfill
\begin{subfigure}[t]{0.48\textwidth}
  \centering
  \includegraphics[width=\linewidth]{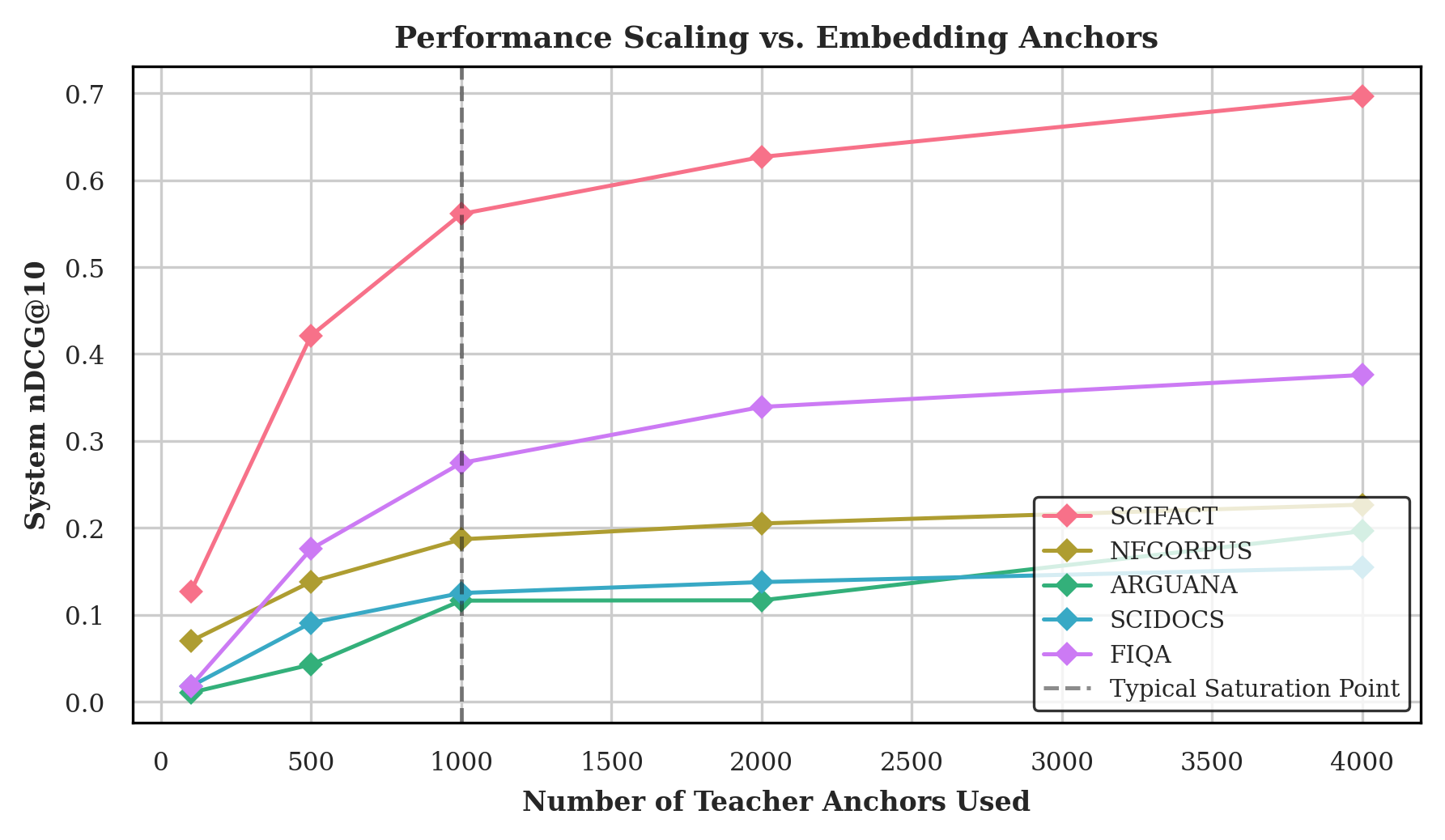}
  \caption{\texttt{text-embedding-004} $\to$ \texttt{gemini-embedding-001}}
\end{subfigure}
\caption{Effect of Cosine FPS anchor count $N$ on retrieval fidelity.
Performance generally improves with more anchors, with diminishing returns on
most datasets.}
\label{fig:anchor_scaling}
\end{figure*}

Table~\ref{tab:data_efficiency} and Figure~\ref{fig:anchor_scaling} evaluate how
many target document embeddings are needed to map $\mathcal{M}_T$ in the Cold
phase. First, Cosine FPS is a loss-independent improvement on clustered corpora:
at $N = 400$, replacing uniform selection with FPS raises T-nDCG@10 not only for
ERA (Arguana: $0.361 \to 0.568$; SciFact: $0.401 \to 0.434$; NFCorpus:
$0.379 \to 0.401$), but also for Embedding-Converter (Arguana: $0.173 \to
0.413$) and LEAF (Arguana: $0.331 \to 0.580$). Second, under Cosine FPS, ERA
outperforms both baselines at every anchor budget on four of five datasets
(SciFact, NFCorpus, SciDocs, FiQA) and stays within $0.012$ of LEAF on Arguana
at $N \geq 400$; at $N = 100$, ERA doubles LEAF's fidelity on FiQA ($0.036$
vs.\ $0.017$) and SciDocs ($0.063$ vs.\ $0.031$). Third, on heavy-tailed
corpora with isolated outliers such as SciDocs (and FiQA at small budgets),
uniform sampling
slightly edges pure FPS because farthest-point selection spends part of its
small budget on peripheral outlier documents. Across both source encoders,
gains diminish as $N$ grows (Figure~\ref{fig:anchor_scaling}), so a few
thousand anchors suffice rather than the full corpus.

\subsection{Progressive Adaptation Across Query Phases}
\label{subsec:phase_progression}

As unlabeled queries become available, ERA transitions from the Cold phase
(document anchors only) to the Independent phase (unlabeled in-domain training
queries) and the Foundation phase (pooled cross-domain English queries). Table~\ref{tab:phase_summary} summarizes this progression at $0\%$ and $20\%$
escalation budgets; full nDCG@10 and T-nDCG@10 matrices across all phases and
budgets are reported in Appendix~\ref{appendix:A}
(Tables~\ref{tab:unified_absolute_performance}--\ref{tab:heterogeneous_fidelity}).
Because queries and documents occupy distinct sub-regions of contrastive
embedding spaces, adding unlabeled queries progressively closes the
query-to-document offset: for \texttt{Qwen-Embedding-3-0.6B} $\to$
\texttt{gemini-embedding-001} at $0\%$ routing, transitioning from Cold to
Foundation lifts SciFact nDCG@10 from $0.676$ to $0.840$, FiQA from $0.289$ to
$0.458$, and Arguana from $0.312$ to $0.559$. Because the escalation threshold
$\gamma$ is calibrated on held-out validation queries without peeking at the
test distribution, the empirical routed fraction naturally varies around the
nominal quantile budget as query distributions shift between validation and test
splits. The deviation is largest for Arguana and SciDocs, which provide no
training queries, so their thresholds are calibrated on held-out anchor
documents rather than queries.

\subsection{Hard-Query Escalation and System Latency}
\label{subsec:routing_eval}

\begin{table}[tb]
\caption{Escalation rate by query disagreement gap (target nDCG minus projected
nDCG) at a 20\% budget across all five BEIR datasets. \textbf{Agnostic} is a
single uniformly random escalation draw at the same routed count as ERA. Because
hard-query buckets are small (e.g., $n{=}8$ on SciFact), a random policy's
capture of them varies widely from draw to draw, whereas ERA's escalation is
deterministic (see text for exact miss probabilities). Here ERA escalates the
20\% of test queries with the lowest router scores (\texttt{text-embedding-004}
$\to$ \texttt{gemini-embedding-001}, Foundation phase).}
\label{tab:router_gaps}
\centering
\small
\setlength{\tabcolsep}{1.5pt}
\begin{tabular*}{0.75\textwidth}{@{\extracolsep{\fill}}llccc@{}}
\toprule
\textbf{Dataset} & \textbf{Gap Bucket} & $|\mathcal{Q}|$ & \textbf{Agnostic} & \textbf{ERA} \\
\midrule
SciFact  & Win ($<0$)         & 11  & 18.2\% & \textbf{9.1\%} \\
         & Small ($0$--$.4$)  & 215 & 21.9\% & 20.5\% \\
         & Large ($.4$--$.6$) & 13  & 7.7\%  & \textbf{15.4\%} \\
         & Huge ($>.6$)       & 8   & 0.0\%  & \textbf{37.5\%} \\
\midrule
SciDocs  & Win ($<0$)         & 142 & 16.9\% & 19.7\% \\
         & Small ($0$--$.4$)  & 782 & 21.2\% & 19.9\% \\
         & Large ($.4$--$.6$) & 45  & 15.6\% & 15.6\% \\
         & Huge ($>.6$)       & 25  & 8.0\%  & \textbf{32.0\%} \\
\midrule
NFCorpus & Win ($<0$)         & 36  & 13.9\% & 13.9\% \\
         & Small ($0$--$.4$)  & 140 & 20.7\% & 20.0\% \\
         & Large ($.4$--$.6$) & 6   & 33.3\% & \textbf{50.0\%} \\
         & Huge ($>.6$)       & 3   & 33.3\% & 33.3\% \\
\midrule
Arguana  & Win ($<0$)         & 194 & 24.2\% & \textbf{19.1\%} \\
         & Small ($0$--$.4$)  & 880 & 18.9\% & 19.4\% \\
         & Large ($.4$--$.6$) & 132 & 19.7\% & \textbf{24.2\%} \\
         & Huge ($>.6$)       & 195 & 21.0\% & 20.5\% \\
\midrule
FiQA     & Win ($<0$)         & 81  & 16.0\% & 23.5\% \\
         & Small ($0$--$.4$)  & 469 & 21.3\% & 21.3\% \\
         & Large ($.4$--$.6$) & 46  & 21.7\% & 13.0\% \\
         & Huge ($>.6$)       & 52  & 13.5\% & 9.6\%  \\
\bottomrule
\end{tabular*}
\end{table}

\begin{table}[tb]
\caption{Component latencies profiled in JAX and against the live target API.
End-to-end ERA latency (\texttt{Qwen-Embedding-3-0.6B} $\to$
\texttt{gemini-embedding-001}, Foundation phase, nominal 20\% budget) is
estimated as local encoding plus adapter/router overhead plus the observed
routed fraction times the target API round trip.}
\label{tab:latency}
\centering
\small
\setlength{\tabcolsep}{2pt}
\begin{tabular*}{0.6\textwidth}{@{\extracolsep{\fill}}llc@{}}
\toprule
\textbf{Stage} & \textbf{Setting} & \textbf{Latency (ms)} \\
\midrule
Adapter MLP & JAX Forward       & $0.62\pm0.04$ \\
Router MLP  & JAX Forward       & $0.58\pm0.03$ \\
Qwen-0.6B   & Local Encoder     & ${\sim}20.00$ \\
Gemini API  & Target API RTT    & $160.45\pm14.98$ \\
\midrule
FiQA        & ERA (20\% Budget) & $37.7$ \\
SciFact     & ERA (20\% Budget) & $55.1$ \\
NFCorpus    & ERA (20\% Budget) & $61.9$ \\
SciDocs     & ERA (20\% Budget) & $82.3$ \\
Arguana     & ERA (20\% Budget) & $85.5$ \\
\bottomrule
\end{tabular*}
\end{table}

\paragraph{Hard-query capture.}
In production retrieval, the majority of traffic consists of routine queries
where the projected embedding already matches the target index, while a small
tail of complex queries accounts for the worst retrieval failures. Because
dataset-wide average nDCG weights routine and failure-mode queries equally,
Table~\ref{tab:router_gaps} stratifies test queries across all five datasets by
their \emph{disagreement gap}---the drop in nDCG@10 when using the adapter
instead of the native target encoder---under a $20\%$ escalation budget. On four of the
five datasets (SciFact, SciDocs, NFCorpus, and Arguana), the ERA router
escalates a higher fraction of high-disagreement queries and/or wastes fewer
escalations on queries where the adapter already matches or beats the target
encoder ($\text{Gap} < 0.0$): on SciFact, ERA captures $37.5\%$ of Huge-Gap
($>0.6$) and $15.4\%$ of Large-Gap ($0.4$--$0.6$) queries compared to $0.0\%$
and $7.7\%$ under difficulty-agnostic allocation, while cutting wasted
escalations in half ($9.1\%$ vs.\ $18.2\%$); on SciDocs, ERA captures $32.0\%$
of Huge-Gap queries vs.\ $8.0\%$; and on NFCorpus, it captures $50.0\%$ of
Large-Gap queries vs.\ $33.3\%$. On noisy financial queries (FiQA), the
lightweight $0.6$B router struggles to separate tail difficulty from surface
query features ($9.6\%$ vs.\ $13.5\%$).

Production deployments observe a single realization of a routing policy, so
the relevant risk of difficulty-agnostic allocation is not its average capture
rate but how often it misses the hardest queries entirely. Sampling the same
routed count uniformly at random (an exact hypergeometric computation), a random
policy escalates \emph{none} of SciFact's eight Huge-Gap queries with
probability $15.9\%$---roughly one deployment in six---and matches ERA's
capture of three or more only $20.6\%$ of the time. On SciDocs, ERA's $32.0\%$
Huge-Gap capture exceeds that of $89\%$ of random draws. ERA's router, by
contrast, is deterministic: the same queries are escalated on every
deployment. We note that these buckets are small, so the router should be
viewed as a safety net for rare failure modes on structured domains rather
than as a calibrated difficulty predictor.

\paragraph{Cross-API routing and hardware latency.}
We also evaluate the heterogeneous API configuration where $f_S$ is a commercial
encoder (\texttt{text-embedding-004}, $768$-D), $f_T$ is
\texttt{gemini-embedding-001} ($3072$-D), and the open-weight Qwen encoder acts
solely as the router $f_R$ so escalated queries never pay for two commercial
APIs (Appendix~\ref{appendix:A},
Tables~\ref{tab:heterogeneous_absolute_performance}--\ref{tab:heterogeneous_fidelity}).
On SciFact at a $20\%$ budget ($20.5\%$ routed), this configuration reaches
$0.868$ nDCG@10 (vs.\ $0.916$ native Gemini), and at an $80\%$ budget on Arguana
it reaches $0.622$ (vs.\ $0.633$). Table~\ref{tab:latency} reports hardware
latency profiled in JAX alongside live target API calls: the local SwiGLU
adapter ($0.62$\,ms) and router ($0.58$\,ms) add only ${\approx}1.2$\,ms of total
overhead per query; with the local \texttt{Qwen-Embedding-3-0.6B} source, this
yields estimated end-to-end query latencies of $37.7$--$85.5$\,ms at a nominal
$20\%$ escalation budget compared to $>160$\,ms when calling the target API on
every query.

\paragraph{Reverse-direction negative control.}
Finally, the reverse direction, which projects the $3072$-D
\texttt{gemini-embedding-001} encoder into the $1024$-D
\texttt{Qwen-Embedding-3-0.6B} index, confirms the dimensionality bottleneck (Appendix~\ref{appendix:A},
Tables~\ref{tab:reverse_absolute_performance}--\ref{tab:reverse_fidelity}):
accuracy remains far below native Gemini on ArguAna ($0.633$), bounded by the
frozen 1024-D document index, while the pure adapter (0\% routing) reaches
$0.78$--$0.80$ T-nDCG@10 fidelity to the Qwen index, indicating that ERA
reproduces the target index geometry rather than acting as a generic score
regularizer.

\section{Conclusion}
\label{sec:conclusion}

We introduced ERA (Embed, Route, Adapt), an end-to-end system for cold-start
cross-model latent retrieval that enables queries encoded by one model to
directly search an existing vector index built by another, without re-embedding
the corpus and without human relevance labels. By selecting anchor documents via
Cosine Farthest Point Sampling, training a lightweight adapter with a Manifold
Alignment Loss ($\mathcal{L}_{\cos} + \mathcal{L}_{\mathrm{GOR}}$) that prevents
representation collapse in the data-scarce regime, and pairing the adapter with
a deterministic router that escalates the rare, complex queries that govern
perceived quality, ERA matches or outperforms point-wise $\ell_2$ and
embedding-conversion baselines across five BEIR benchmarks while adding only
${\approx}1.2$\,ms of adapter and router compute beyond the local encoder.

\section*{Limitations}

Our study highlights four practical boundaries that motivate future work:
\begin{itemize}
\item \textbf{Dimensionality ceiling:} When $d_S < d_T$, an asymmetric adapter
cannot recover semantic distinctions that the source encoder never captured; as
confirmed by our reverse-direction negative control (Gemini $\to$ Qwen), ERA
faithfully reproduces the target manifold's geometry up to its native capacity
ceiling.
\item \textbf{Outlier sensitivity of pure FPS:} On heavy-tailed corpora with
disconnected outlier documents (such as SciDocs), greedy Cosine FPS can spend
part of a small anchor budget on peripheral outliers rather than the dense
semantic core. Exploring quantile-based farthest point sampling (e.g., selecting
the $q$-th percentile farthest point) is a promising direction.
\item \textbf{Router discrimination on noisy queries:} While the lightweight
$0.6$B router reliably isolates large-gap failure queries on structured domains
(SciFact, SciDocs, NFCorpus, Arguana), it behaves closer to uniform load
balancing on short, noisy financial queries (FiQA), and static validation
quantiles can drift under test-time distribution shifts.
\item \textbf{Language and scale scope:} Our empirical evaluation focuses on
English retrieval benchmarks; extending cross-model latent projection and routing
to multilingual retrieval (e.g., MIRACL) and larger-scale corpora is for future work.
\end{itemize}

\section*{AI Assistance}
An AI assistant was used to help draft and edit text and to format LaTeX
tables and the bibliography. All experiments, results, and claims were
produced and verified by the authors.

\bibliographystyle{abbrvnat}
\bibliography{custom}

%%%%%%%%%%%%%%%%%%%%%%%%%%%%%%%%%%%%%%%%%%%%%%%%%%%%%%%%%%%%
\appendix
\newpage
\section{Extended Empirical Results}
\label{appendix:A}

This appendix provides the complete system accuracy (nDCG@10) and topological
fidelity (T-nDCG@10) matrices for the \texttt{Qwen-Embedding-3-0.6B} $\to$
\texttt{gemini-embedding-001} pipeline
(Tables~\ref{tab:unified_absolute_performance} and \ref{tab:unified_fidelity}),
the full accuracy and fidelity matrices for the heterogeneous API configuration
(\texttt{text-embedding-004} $\to$ \texttt{gemini-embedding-001} routed by Qwen;
Tables~\ref{tab:heterogeneous_absolute_performance} and
\ref{tab:heterogeneous_fidelity}), and the reverse-direction negative control
(\texttt{gemini-embedding-001} $\to$ \texttt{Qwen-Embedding-3-0.6B};
Tables~\ref{tab:reverse_absolute_performance} and \ref{tab:reverse_fidelity}).

\begin{table*}[h!]
\caption{System nDCG@10 across adaptation phases and escalation budgets for
\texttt{Qwen-Embedding-3-0.6B} $\to$ \texttt{gemini-embedding-001}. The
\textbf{Routed} column reports the observed fraction of test queries escalated
to the target encoder.}
\label{tab:unified_absolute_performance}
\centering
\small
\setlength{\tabcolsep}{4pt}
\resizebox{\textwidth}{!}{%
\begin{tabular}{@{}llccccc@{}}
\toprule
\textbf{Dataset} & \textbf{Budget} & \textbf{Routed (Cold $\mid$ Ind. $\mid$ Found.)} & \textbf{Cold} & \textbf{Independent} & \textbf{Foundation} & \textbf{Target (Gemini)} \\
\midrule
SciFact
& 0\%  & 0.0\% $\mid$ 0.0\% $\mid$ 0.0\%    & 0.676$\pm$0.006 & 0.837$\pm$0.003 & \textbf{0.840$\pm$0.006} & 0.916$\pm$0.000 \\
& 20\% & 20.6\% $\mid$ 20.8\% $\mid$ 21.1\% & 0.733$\pm$0.004 & 0.857$\pm$0.004 & \textbf{0.860$\pm$0.001} & \\
& 50\% & 42.2\% $\mid$ 40.5\% $\mid$ 41.8\% & 0.780$\pm$0.007 & 0.872$\pm$0.002 & \textbf{0.876$\pm$0.005} & \\
& 80\% & 86.0\% $\mid$ 86.1\% $\mid$ 87.0\% & 0.875$\pm$0.002 & 0.898$\pm$0.000 & \textbf{0.901$\pm$0.001} & \\
\midrule
FiQA
& 0\%  & 0.0\% $\mid$ 0.0\% $\mid$ 0.0\%    & 0.289$\pm$0.000 & 0.354$\pm$0.003 & \textbf{0.458$\pm$0.004} & 0.620$\pm$0.000 \\
& 20\% & 10.5\% $\mid$ 9.7\% $\mid$ 10.3\%  & 0.325$\pm$0.004 & 0.376$\pm$0.003 & \textbf{0.473$\pm$0.004} & \\
& 50\% & 43.0\% $\mid$ 45.0\% $\mid$ 45.4\% & 0.429$\pm$0.004 & 0.468$\pm$0.002 & \textbf{0.518$\pm$0.007} & \\
& 80\% & 79.6\% $\mid$ 81.2\% $\mid$ 81.2\% & 0.548$\pm$0.003 & 0.566$\pm$0.002 & \textbf{0.575$\pm$0.008} & \\
\midrule
NFCorpus
& 0\%  & 0.0\% $\mid$ 0.0\% $\mid$ 0.0\%    & 0.211$\pm$0.004 & 0.224$\pm$0.006 & \textbf{0.228$\pm$0.008} & 0.276$\pm$0.000 \\
& 20\% & 24.9\% $\mid$ 25.0\% $\mid$ 25.4\% & 0.231$\pm$0.003 & 0.241$\pm$0.006 & \textbf{0.247$\pm$0.007} & \\
& 50\% & 51.2\% $\mid$ 51.5\% $\mid$ 50.3\% & 0.258$\pm$0.000 & 0.260$\pm$0.001 & \textbf{0.263$\pm$0.001} & \\
& 80\% & 73.0\% $\mid$ 73.2\% $\mid$ 73.3\% & 0.264$\pm$0.000 & 0.266$\pm$0.002 & \textbf{0.266$\pm$0.001} & \\
\midrule
Arguana
& 0\%  & 0.0\% $\mid$ 0.0\% $\mid$ 0.0\%    & 0.312$\pm$0.003 & 0.306$\pm$0.006 & \textbf{0.559$\pm$0.007} & 0.633$\pm$0.000 \\
& 20\% & 40.4\% $\mid$ 40.0\% $\mid$ 40.1\% & 0.449$\pm$0.001 & 0.443$\pm$0.002 & \textbf{0.595$\pm$0.005} & \\
& 50\% & 71.0\% $\mid$ 70.6\% $\mid$ 71.2\% & 0.553$\pm$0.003 & 0.551$\pm$0.002 & \textbf{0.619$\pm$0.004} & \\
& 80\% & 93.3\% $\mid$ 93.1\% $\mid$ 93.5\% & 0.619$\pm$0.000 & 0.619$\pm$0.001 & \textbf{0.631$\pm$0.001} & \\
\midrule
SciDocs
& 0\%  & 0.0\% $\mid$ 0.0\% $\mid$ 0.0\%    & 0.173$\pm$0.001 & 0.175$\pm$0.001 & \textbf{0.180$\pm$0.002} & 0.253$\pm$0.000 \\
& 20\% & 38.3\% $\mid$ 38.3\% $\mid$ 38.1\% & 0.203$\pm$0.001 & 0.205$\pm$0.001 & \textbf{0.207$\pm$0.002} & \\
& 50\% & 72.6\% $\mid$ 73.8\% $\mid$ 74.0\% & 0.228$\pm$0.001 & 0.228$\pm$0.000 & \textbf{0.228$\pm$0.001} & \\
& 80\% & 92.5\% $\mid$ 92.9\% $\mid$ 93.0\% & 0.246$\pm$0.000 & 0.246$\pm$0.000 & \textbf{0.245$\pm$0.000} & \\
\bottomrule
\end{tabular}}
\end{table*}

\begin{table*}[t]
\caption{Topological fidelity to the target index (T-nDCG@10, treating target
encoder top-10 retrieval as ground truth) for \texttt{Qwen-Embedding-3-0.6B}
$\to$ \texttt{gemini-embedding-001}. The final column reports exact top-10
intersection overlap (Recall@10) in the Foundation phase.}
\label{tab:unified_fidelity}
\centering
\small
\setlength{\tabcolsep}{5pt}
\begin{tabular*}{\textwidth}{@{\extracolsep{\fill}}llcccc@{}}
\toprule
\textbf{Dataset} & \textbf{Budget} & \textbf{Cold} & \textbf{Independent} & \textbf{Foundation} & \textbf{Found.\ Recall@10} \\
\midrule
SciFact
& 0\%  & 0.594$\pm$0.003 & 0.728$\pm$0.000 & \textbf{0.736$\pm$0.002} & 67.2\%$\pm$0.3\% \\
& 20\% & 0.687$\pm$0.003 & 0.795$\pm$0.001 & \textbf{0.802$\pm$0.002} & 75.1\%$\pm$0.3\% \\
& 50\% & 0.779$\pm$0.010 & 0.851$\pm$0.011 & \textbf{0.859$\pm$0.001} & 82.2\%$\pm$0.3\% \\
& 80\% & 0.943$\pm$0.001 & 0.959$\pm$0.002 & \textbf{0.961$\pm$0.002} & 95.4\%$\pm$0.4\% \\
\midrule
FiQA
& 0\%  & 0.395$\pm$0.001 & 0.489$\pm$0.001 & \textbf{0.493$\pm$0.001} & 43.1\%$\pm$0.3\% \\
& 20\% & 0.471$\pm$0.009 & 0.546$\pm$0.008 & \textbf{0.555$\pm$0.005} & 49.9\%$\pm$0.7\% \\
& 50\% & 0.669$\pm$0.007 & 0.731$\pm$0.009 & \textbf{0.736$\pm$0.004} & 70.2\%$\pm$0.5\% \\
& 80\% & 0.884$\pm$0.004 & 0.911$\pm$0.003 & \textbf{0.911$\pm$0.001} & 90.0\%$\pm$0.2\% \\
\midrule
NFCorpus
& 0\%  & 0.532$\pm$0.001 & \textbf{0.590$\pm$0.002} & 0.589$\pm$0.005 & 52.5\%$\pm$0.5\% \\
& 20\% & 0.691$\pm$0.003 & 0.734$\pm$0.003 & \textbf{0.736$\pm$0.006} & 68.4\%$\pm$0.7\% \\
& 50\% & 0.818$\pm$0.000 & \textbf{0.836$\pm$0.000} & 0.834$\pm$0.008 & 80.1\%$\pm$0.8\% \\
& 80\% & 0.909$\pm$0.010 & 0.918$\pm$0.004 & \textbf{0.920$\pm$0.001} & 90.2\%$\pm$0.1\% \\
\midrule
Arguana
& 0\%  & 0.666$\pm$0.002 & 0.661$\pm$0.004 & \textbf{0.762$\pm$0.001} & 70.1\%$\pm$0.3\% \\
& 20\% & 0.809$\pm$0.003 & 0.805$\pm$0.003 & \textbf{0.862$\pm$0.000} & 82.7\%$\pm$0.1\% \\
& 50\% & 0.915$\pm$0.001 & 0.913$\pm$0.001 & \textbf{0.938$\pm$0.001} & 92.1\%$\pm$0.2\% \\
& 80\% & 0.980$\pm$0.000 & 0.979$\pm$0.000 & \textbf{0.986$\pm$0.000} & 98.2\%$\pm$0.1\% \\
\midrule
SciDocs
& 0\%  & 0.514$\pm$0.004 & 0.514$\pm$0.002 & \textbf{0.528$\pm$0.004} & 46.9\%$\pm$0.2\% \\
& 20\% & 0.704$\pm$0.002 & 0.706$\pm$0.001 & \textbf{0.715$\pm$0.003} & 67.7\%$\pm$0.2\% \\
& 50\% & 0.872$\pm$0.001 & 0.879$\pm$0.001 & \textbf{0.884$\pm$0.004} & 86.8\%$\pm$0.4\% \\
& 80\% & 0.966$\pm$0.001 & 0.967$\pm$0.001 & \textbf{0.969$\pm$0.001} & 96.5\%$\pm$0.1\% \\
\bottomrule
\end{tabular*}
\end{table*}

\begin{table*}[t]
\caption{System nDCG@10 for the heterogeneous API configuration (router:
\texttt{Qwen-Embedding-3-0.6B}; source query encoder:
\texttt{text-embedding-004}; target index: \texttt{gemini-embedding-001}).
Arguana and SciDocs provide no training queries, so their Independent phase
reduces to Cold.}
\label{tab:heterogeneous_absolute_performance}
\centering
\small
\setlength{\tabcolsep}{4pt}
\resizebox{\textwidth}{!}{%
\begin{tabular}{@{}llccccc@{}}
\toprule
\textbf{Dataset} & \textbf{Budget} & \textbf{Routed (Cold $\mid$ Ind. $\mid$ Found.)} & \textbf{Cold} & \textbf{Independent} & \textbf{Foundation} & \textbf{Target (Gemini)} \\
\midrule
SciFact
& 0\%  & 0.0\% $\mid$ 0.0\% $\mid$ 0.0\%    & 0.700$\pm$0.006 & 0.850$\pm$0.001 & \textbf{0.858$\pm$0.005} & 0.916$\pm$0.000 \\
& 20\% & 19.6\% $\mid$ 20.1\% $\mid$ 20.5\% & 0.751$\pm$0.005 & 0.864$\pm$0.003 & \textbf{0.868$\pm$0.005} & \\
& 50\% & 57.5\% $\mid$ 57.4\% $\mid$ 57.4\% & 0.837$\pm$0.001 & 0.889$\pm$0.001 & \textbf{0.891$\pm$0.004} & \\
& 80\% & 79.6\% $\mid$ 80.4\% $\mid$ 79.2\% & 0.862$\pm$0.003 & 0.894$\pm$0.002 & \textbf{0.898$\pm$0.001} & \\
\midrule
FiQA
& 0\%  & 0.0\% $\mid$ 0.0\% $\mid$ 0.0\%    & 0.378$\pm$0.005 & \textbf{0.464$\pm$0.000} & 0.459$\pm$0.002 & 0.621$\pm$0.000 \\
& 20\% & 11.7\% $\mid$ 11.8\% $\mid$ 12.0\% & 0.402$\pm$0.004 & \textbf{0.475$\pm$0.003} & 0.472$\pm$0.001 & \\
& 50\% & 41.9\% $\mid$ 41.4\% $\mid$ 41.6\% & 0.474$\pm$0.006 & \textbf{0.522$\pm$0.006} & 0.521$\pm$0.003 & \\
& 80\% & 77.4\% $\mid$ 76.7\% $\mid$ 77.5\% & 0.561$\pm$0.004 & 0.579$\pm$0.002 & \textbf{0.580$\pm$0.002} & \\
\midrule
NFCorpus
& 0\%  & 0.0\% $\mid$ 0.0\% $\mid$ 0.0\%    & 0.230$\pm$0.006 & 0.237$\pm$0.004 & \textbf{0.238$\pm$0.004} & 0.277$\pm$0.000 \\
& 20\% & 28.1\% $\mid$ 27.7\% $\mid$ 29.4\% & 0.250$\pm$0.005 & 0.252$\pm$0.004 & \textbf{0.255$\pm$0.004} & \\
& 50\% & 57.2\% $\mid$ 54.8\% $\mid$ 53.5\% & 0.261$\pm$0.001 & 0.262$\pm$0.001 & \textbf{0.264$\pm$0.002} & \\
& 80\% & 89.6\% $\mid$ 89.0\% $\mid$ 89.4\% & \textbf{0.274$\pm$0.001} & 0.273$\pm$0.002 & \textbf{0.274$\pm$0.001} & \\
\midrule
Arguana
& 0\%  & 0.0\% $\mid$ 0.0\% $\mid$ 0.0\%    & 0.197$\pm$0.024 & 0.197$\pm$0.024 & \textbf{0.451$\pm$0.020} & 0.633$\pm$0.000 \\
& 20\% & 41.9\% $\mid$ 41.9\% $\mid$ 42.0\% & 0.383$\pm$0.014 & 0.383$\pm$0.014 & \textbf{0.531$\pm$0.013} & \\
& 50\% & 73.6\% $\mid$ 73.6\% $\mid$ 73.9\% & 0.523$\pm$0.008 & 0.523$\pm$0.008 & \textbf{0.589$\pm$0.008} & \\
& 80\% & 93.0\% $\mid$ 93.0\% $\mid$ 93.5\% & 0.606$\pm$0.002 & 0.606$\pm$0.002 & \textbf{0.622$\pm$0.003} & \\
\midrule
SciDocs
& 0\%  & 0.0\% $\mid$ 0.0\% $\mid$ 0.0\%    & 0.153$\pm$0.001 & 0.153$\pm$0.001 & \textbf{0.161$\pm$0.003} & 0.253$\pm$0.000 \\
& 20\% & 34.4\% $\mid$ 34.3\% $\mid$ 34.5\% & 0.191$\pm$0.002 & 0.191$\pm$0.002 & \textbf{0.196$\pm$0.001} & \\
& 50\% & 70.5\% $\mid$ 70.4\% $\mid$ 70.2\% & 0.225$\pm$0.001 & 0.225$\pm$0.001 & \textbf{0.226$\pm$0.002} & \\
& 80\% & 91.7\% $\mid$ 91.8\% $\mid$ 91.7\% & \textbf{0.247$\pm$0.000} & \textbf{0.247$\pm$0.000} & \textbf{0.247$\pm$0.001} & \\
\bottomrule
\end{tabular}}
\end{table*}

\begin{table*}[t]
\caption{Topological fidelity (T-nDCG@10) and Foundation Recall@10 for the
heterogeneous API configuration (\texttt{text-embedding-004} $\to$
\texttt{gemini-embedding-001}, routed by \texttt{Qwen-Embedding-3-0.6B}).
Arguana and SciDocs provide no training queries, so their Independent phase
reduces to Cold.}
\label{tab:heterogeneous_fidelity}
\centering
\small
\setlength{\tabcolsep}{5pt}
\begin{tabular*}{\textwidth}{@{\extracolsep{\fill}}llcccc@{}}
\toprule
\textbf{Dataset} & \textbf{Budget} & \textbf{Cold} & \textbf{Independent} & \textbf{Foundation} & \textbf{Found.\ Recall@10} \\
\midrule
SciFact
& 0\%  & 0.633$\pm$0.003 & 0.755$\pm$0.002 & \textbf{0.761$\pm$0.001} & 69.3\%$\pm$0.4\% \\
& 20\% & 0.716$\pm$0.004 & 0.813$\pm$0.002 & \textbf{0.819$\pm$0.002} & 76.5\%$\pm$0.4\% \\
& 50\% & 0.856$\pm$0.002 & 0.905$\pm$0.002 & \textbf{0.907$\pm$0.003} & 88.0\%$\pm$0.5\% \\
& 80\% & 0.931$\pm$0.004 & \textbf{0.953$\pm$0.004} & 0.951$\pm$0.006 & 93.7\%$\pm$0.8\% \\
\midrule
FiQA
& 0\%  & 0.489$\pm$0.008 & \textbf{0.604$\pm$0.004} & 0.597$\pm$0.003 & 52.2\%$\pm$0.3\% \\
& 20\% & 0.550$\pm$0.007 & \textbf{0.652$\pm$0.002} & 0.648$\pm$0.003 & 58.2\%$\pm$0.4\% \\
& 50\% & 0.709$\pm$0.009 & \textbf{0.772$\pm$0.008} & 0.769$\pm$0.007 & 72.5\%$\pm$0.8\% \\
& 80\% & 0.890$\pm$0.005 & 0.912$\pm$0.005 & \textbf{0.916$\pm$0.004} & 89.9\%$\pm$0.5\% \\
\midrule
NFCorpus
& 0\%  & 0.610$\pm$0.007 & 0.653$\pm$0.003 & \textbf{0.654$\pm$0.002} & 58.5\%$\pm$0.5\% \\
& 20\% & 0.752$\pm$0.012 & 0.779$\pm$0.005 & \textbf{0.787$\pm$0.011} & 73.7\%$\pm$1.4\% \\
& 50\% & 0.867$\pm$0.006 & \textbf{0.877$\pm$0.001} & 0.872$\pm$0.004 & 84.0\%$\pm$0.3\% \\
& 80\% & 0.969$\pm$0.003 & \textbf{0.974$\pm$0.002} & 0.974$\pm$0.003 & 96.7\%$\pm$0.3\% \\
\midrule
Arguana
& 0\%  & 0.498$\pm$0.029 & 0.498$\pm$0.029 & \textbf{0.714$\pm$0.005} & 64.3\%$\pm$0.5\% \\
& 20\% & 0.716$\pm$0.017 & 0.716$\pm$0.017 & \textbf{0.840$\pm$0.003} & 79.9\%$\pm$0.3\% \\
& 50\% & 0.872$\pm$0.009 & 0.872$\pm$0.009 & \textbf{0.929$\pm$0.003} & 91.1\%$\pm$0.2\% \\
& 80\% & 0.967$\pm$0.003 & 0.967$\pm$0.003 & \textbf{0.983$\pm$0.000} & 97.9\%$\pm$0.0\% \\
\midrule
SciDocs
& 0\%  & 0.500$\pm$0.003 & 0.500$\pm$0.003 & \textbf{0.521$\pm$0.003} & 46.2\%$\pm$0.4\% \\
& 20\% & 0.682$\pm$0.002 & 0.682$\pm$0.002 & \textbf{0.694$\pm$0.006} & 65.5\%$\pm$0.7\% \\
& 50\% & 0.864$\pm$0.003 & 0.864$\pm$0.003 & \textbf{0.867$\pm$0.001} & 84.9\%$\pm$0.3\% \\
& 80\% & 0.961$\pm$0.001 & 0.961$\pm$0.001 & \textbf{0.962$\pm$0.001} & 95.7\%$\pm$0.1\% \\
\bottomrule
\end{tabular*}
\end{table*}

\begin{table*}[t]
\caption{Reverse-direction negative control (\texttt{gemini-embedding-001}
$\to$ \texttt{Qwen-Embedding-3-0.6B}): ground-truth retrieval accuracy
(nDCG@10). Projecting a $3072$-D encoder into a $1024$-D target index is bounded
by the lower-dimensional target document index.}
\label{tab:reverse_absolute_performance}
\centering
\small
\setlength{\tabcolsep}{4pt}
\resizebox{\textwidth}{!}{%
\begin{tabular}{@{}llcccc@{}}
\toprule
\textbf{Dataset (Native Gemini)} & \textbf{Budget} & \textbf{Routed (Ind. $\mid$ Found.)} & \textbf{Independent} & \textbf{Foundation} & \textbf{Target (Qwen)} \\
\midrule
SciFact (0.916)
& 0\%  & 0.0\% $\mid$ 0.0\%   & 0.761$\pm$0.003 & \textbf{0.765$\pm$0.003} & 0.726$\pm$0.000 \\
& 20\% & 26.9\% $\mid$ 26.2\% & 0.753$\pm$0.004 & \textbf{0.755$\pm$0.001} & \\
& 50\% & 51.8\% $\mid$ 54.5\% & 0.747$\pm$0.002 & \textbf{0.751$\pm$0.003} & \\
& 80\% & 81.1\% $\mid$ 82.3\% & 0.736$\pm$0.002 & \textbf{0.738$\pm$0.002} & \\
\midrule
FiQA (0.621)
& 0\%  & 0.0\% $\mid$ 0.0\%   & 0.378$\pm$0.004 & \textbf{0.400$\pm$0.001} & 0.469$\pm$0.000 \\
& 20\% & 20.7\% $\mid$ 19.8\% & 0.403$\pm$0.002 & \textbf{0.419$\pm$0.002} & \\
& 50\% & 50.1\% $\mid$ 48.5\% & 0.436$\pm$0.001 & \textbf{0.442$\pm$0.002} & \\
& 80\% & 78.2\% $\mid$ 78.9\% & 0.450$\pm$0.000 & \textbf{0.454$\pm$0.001} & \\
\midrule
NFCorpus (0.277)
& 0\%  & 0.0\% $\mid$ 0.0\%   & \textbf{0.238$\pm$0.005} & 0.231$\pm$0.002 & 0.242$\pm$0.000 \\
& 20\% & 5.2\% $\mid$ 6.3\%   & \textbf{0.240$\pm$0.004} & 0.233$\pm$0.001 & \\
& 50\% & 33.0\% $\mid$ 33.0\% & \textbf{0.237$\pm$0.002} & 0.234$\pm$0.002 & \\
& 80\% & 72.1\% $\mid$ 71.4\% & \textbf{0.241$\pm$0.002} & 0.239$\pm$0.003 & \\
\midrule
Arguana (0.633)
& 0\%  & 0.0\% $\mid$ 0.0\%   & \textbf{0.509$\pm$0.003} & 0.480$\pm$0.004 & 0.486$\pm$0.000 \\
& 20\% & 90.8\% $\mid$ 90.8\% & \textbf{0.489$\pm$0.002} & 0.487$\pm$0.000 & \\
& 50\% & 98.9\% $\mid$ 98.9\% & \textbf{0.487$\pm$0.000} & 0.486$\pm$0.000 & \\
& 80\% & 99.8\% $\mid$ 99.8\% & \textbf{0.486$\pm$0.000} & 0.486$\pm$0.000 & \\
\midrule
SciDocs (0.253)
& 0\%  & 0.0\% $\mid$ 0.0\%   & 0.195$\pm$0.001 & \textbf{0.205$\pm$0.003} & 0.229$\pm$0.000 \\
& 20\% & 95.7\% $\mid$ 95.8\% & 0.226$\pm$0.001 & \textbf{0.227$\pm$0.001} & \\
& 50\% & 99.6\% $\mid$ 99.6\% & \textbf{0.228$\pm$0.000} & \textbf{0.228$\pm$0.000} & \\
& 80\% & 99.9\% $\mid$ 99.9\% & \textbf{0.228$\pm$0.000} & \textbf{0.228$\pm$0.000} & \\
\bottomrule
\end{tabular}}
\end{table*}

\begin{table*}[t]
\caption{Reverse-direction negative control (\texttt{gemini-embedding-001}
$\to$ \texttt{Qwen-Embedding-3-0.6B}): topological fidelity to the target Qwen
index (T-nDCG@10) and Foundation Recall@10.}
\label{tab:reverse_fidelity}
\centering
\small
\setlength{\tabcolsep}{6pt}
\begin{tabular*}{\textwidth}{@{\extracolsep{\fill}}llccc@{}}
\toprule
\textbf{Dataset} & \textbf{Budget} & \textbf{Independent} & \textbf{Foundation} & \textbf{Found.\ Recall@10} \\
\midrule
SciFact
& 0\%  & 0.738$\pm$0.002 & \textbf{0.746$\pm$0.005} & 67.1\%$\pm$0.6\% \\
& 20\% & 0.816$\pm$0.003 & \textbf{0.822$\pm$0.009} & 76.7\%$\pm$1.2\% \\
& 50\% & 0.877$\pm$0.004 & \textbf{0.887$\pm$0.010} & 85.2\%$\pm$1.2\% \\
& 80\% & 0.953$\pm$0.003 & \textbf{0.958$\pm$0.001} & 94.5\%$\pm$0.1\% \\
\midrule
FiQA
& 0\%  & 0.542$\pm$0.001 & \textbf{0.566$\pm$0.002} & 49.2\%$\pm$0.1\% \\
& 20\% & 0.637$\pm$0.003 & \textbf{0.653$\pm$0.001} & 59.5\%$\pm$0.2\% \\
& 50\% & 0.772$\pm$0.003 & \textbf{0.777$\pm$0.006} & 73.9\%$\pm$0.6\% \\
& 80\% & 0.900$\pm$0.006 & \textbf{0.908$\pm$0.002} & 89.2\%$\pm$0.3\% \\
\midrule
NFCorpus
& 0\%  & 0.626$\pm$0.004 & \textbf{0.635$\pm$0.004} & 56.7\%$\pm$0.2\% \\
& 20\% & 0.641$\pm$0.008 & \textbf{0.650$\pm$0.001} & 58.7\%$\pm$0.4\% \\
& 50\% & 0.752$\pm$0.004 & \textbf{0.752$\pm$0.010} & 70.8\%$\pm$1.1\% \\
& 80\% & \textbf{0.893$\pm$0.006} & 0.893$\pm$0.009 & 87.2\%$\pm$1.0\% \\
\midrule
Arguana
& 0\%  & 0.780$\pm$0.002 & \textbf{0.798$\pm$0.001} & 73.0\%$\pm$0.1\% \\
& 20\% & 0.982$\pm$0.001 & \textbf{0.985$\pm$0.001} & 98.0\%$\pm$0.1\% \\
& 50\% & 0.997$\pm$0.000 & \textbf{0.998$\pm$0.000} & 99.7\%$\pm$0.0\% \\
& 80\% & 0.999$\pm$0.000 & \textbf{1.000$\pm$0.000} & 99.9\%$\pm$0.0\% \\
\midrule
SciDocs
& 0\%  & 0.563$\pm$0.003 & \textbf{0.590$\pm$0.002} & 49.7\%$\pm$0.3\% \\
& 20\% & 0.980$\pm$0.003 & \textbf{0.982$\pm$0.004} & 98.0\%$\pm$0.3\% \\
& 50\% & \textbf{0.998$\pm$0.000} & \textbf{0.998$\pm$0.000} & 99.8\%$\pm$0.0\% \\
& 80\% & \textbf{0.999$\pm$0.000} & 0.999$\pm$0.000 & 99.9\%$\pm$0.0\% \\
\bottomrule
\end{tabular*}
\end{table*}

%%%%%%%%%%%%%%%%%%%%%%%%%%%%%%%%%%%%%%%%%%%%%%%%%%%%%%%%%%%%

\end{document}